\documentclass[%
 reprint,
superscriptaddress,
 amsmath,amssymb,
 prd,
]{revtex4-2}

\usepackage{graphicx}%
\usepackage{dcolumn}%
\usepackage{bm}%

\usepackage{xspace}
\usepackage[normalem]{ulem} %
\usepackage{color}
\usepackage{natbib}
\usepackage[T1]{fontenc}
\usepackage{orcidlink}

\begin{document}

\preprint{APS/123-QED}

\title{Modeling Compact Binary Merger Waveforms Beyond General Relativity}%

\author{Gabriel S.~Bonilla \orcidlink{0000-0003-4502-528X}}
\email{gsb76@cornell.edu}
\affiliation{Cornell Center for Astrophysics and Planetary Science, Cornell
University, Ithaca, New York 14853, USA}
\affiliation{Nicholas and Lee Begovich Center for Gravitational-Wave Physics and Astronomy,
California State University, Fullerton, Fullerton, California 92831, USA}
\author{Prayush Kumar \orcidlink{0000-0001-5523-4603}}
\affiliation{Cornell Center for Astrophysics and Planetary Science, Cornell
University, Ithaca, New York 14853, USA}
\affiliation{International Centre for Theoretical Sciences, Tata Institute of Fundamental Research, Bangalore 560089, India}
\author{Saul A.~Teukolsky \orcidlink{0000-0001-9765-4526}}
\affiliation{Cornell Center for Astrophysics and Planetary Science, Cornell
University, Ithaca, New York 14853, USA}
\affiliation{Theoretical Astrophysics 350-17,
        California Institute of Technology, Pasadena, CA 91125, USA}

\date{\today}%

\begin{abstract}
The parameterized post-Einsteinian framework modifies inspiral waveform models to incorporate effects beyond General Relativity. We extend the existing model into the merger-ringdown regime. The modification introduced here adds a single degree of freedom that corresponds to a change in the binary coalescence time. Other merger properties remain as predicted by GR. We discuss parameter estimation with this model, and how it can be used to extract information from beyond-GR waveforms.
\end{abstract}

\maketitle

\section{\label{sec:Introduction}Introduction}

For over two hundred years Newton's theory of gravity stood without equal as an accurate description of gravity. For the last hundred years, general relativity has superseded Newtonian gravity in predictive power. However, there is good reason
to believe that general relativity (GR) is not the ultimate theory of gravity.
In particular, many theorists believe that in the classical limit of some future quantum theory of gravity, there will be modifications that may be detectable on classical length scales ~\cite{Berti2018-su, PhysRevD.94.084002, LIGOScientific:2016lio}.

GR reduces to Newtonian gravity in the slow-motion, weak-field limit, but it is not merely a correction term that is added on top of Newtonian predictions. Full GR is only obtained by demanding one's model of gravity adhere to new principles not present in Newtonian gravity, often referred to as the pillars of GR~\cite{Berti2018-su}. As a result, GR predicts additional phenomena that are entirely absent in the Newtonian theory, such as black holes and gravitational waves(GWs). Likewise, we expect that the theory of gravity to replace GR will be founded on new principles and similarly reveal new phenomena not present in GR. It is possible that evidence of beyond-GR phenomena will make their first observable appearance as small deviations from GR in these GWs~\cite{PhysRevD.96.084039}.

The detection of GW events such as GW150914 and many others~\cite{PhysRevX.9.031040,PhysRevX.11.021053,LIGOScientific:2016lio,PhysRevD.100.104036,LIGOScientific:2021djp} since then have provided direct evidence that gravitational waves, not present in Newtonian gravity, are present in our universe and that GR is an accurate description of gravity, even in the strong-field regime. These detections are made possible by searching for signals in the data of GW detectors such as LIGO using our knowledge of what we already expect these waveforms to look like in GR~\cite{PhysRevX.9.031040,PhysRevX.11.021053,LIGOScientific:2016aoc, LIGOScientific:2021usb, LIGOScientific:2021djp}. The accurate construction of these model waveforms, used as matched-filter templates, is being improved upon as the sensitivity of the GW detectors improves.

For the process of matched-filtering to work however, the signals in GW data must be sufficiently similar to the constructed templates. By limiting oneself to only using GR to generate templates then, it can become difficult to even detect a waveform that deviates from GR. For loud enough signals, though, a binary coalescence with certain parameters in a beyond-GR theory may be closely mimicked by a coalescence model in GR with slightly different parameters~\cite{PhysRevD.78.124020}.
The choice of using GR in the construction of waveform templates then, by construction, limits our ability to detect deviations from GR. Assuming GR for binaries where GR may be violated may cause us to ascribe incorrect properties to these binaries, and this will lead us to draw incorrect conclusions about the astrophysical population these binaries come from~\cite{Berti2018-su}.

There exist alternative theories of gravity beyond GR such as dynamical Chern-Simons gravity and Einstein dilaton Gauss Bonnet where there has been partial progress in producing GW waveforms~\cite{Okounkova2020-nf}. Even if it were possible to produce waveforms from these beyond-GR theories with the full technology that we have for GR waveforms, it would still be computationally prohibitive to test GW data against each beyond-GR theory individually. It might also be the case that some yet-unknown beyond-GR theory fits the signal in our data even better than any of the ones being used in the parameter estimation. 

Yunes and Pretorius have introduced a framework they refer to as the \emph{parameterized post-Einsteinian}(ppE) formalism~\cite{PhysRevD.80.122003} for modifying waveforms such that neither GR nor any other particular beyond-GR theory is assumed, so long as the beyond-GR theory is smoothly connected to GR. Although it is not yet possible to perform full numerical simulations of mergers in all beyond-GR theories, it is possible in many to calculate the leading-PN order corrections to the waveform~\cite{PhysRevD.98.084042}. This correction is then incorporated via a modification of the corresponding waveform in GR. This allows for the searching of evidence for beyond-GR theories using GW data in a model-independent manner.

A limitation of Yunes and Pretorius' model that is addressed in this paper is that their model is valid only during the inspiral regime of the binary coalescence process. The Yunes and Pretorius model parameterizes deviations to the orbital phasing as leading PN-order corrections to the GR waveform in the inspiral regime. This approximation breaks down before the merger is reached. This is relevant to current efforts in GW detection with terrestrial observatories, since a significant portion of the SNR of most commonly detected (higher mass) sources is in the post-inspiral regime, where the in-band waveform is dominated by merger and ringdown. Yunes and Pretorius' model does not extend into this regime and does not model beyond-GR waveforms via corrections to GR there.
In this paper, we present a new extension to the ppE framework that allows for corrections to the GR waveform past the inspiral regime, with a functional form that is agnostic to the underlying GR approximant being used. The new extension decouples the merger portion of the waveform from the initial inspiral assuming a minimal number of new parameters. Through it we are able to model changes to the rate of coalescence in GW signals all the way to merger, independently of the changes that occur in the inspiral regime.
Our extended-PPE framework is capable of measuring beyond-GR effects in signals coming from heavy black hole binaries that LIGO-Virgo detectors have been observing as their most common GW events~\cite{PhysRevX.9.031040, PhysRevX.11.021053, LIGOScientific:2016aoc, LIGOScientific:2017bnn}.

Rest of this paper is organized as follows. In Sec.~\ref{sec:grapprox}, we discuss the existing gravitational waveform models used, in particular the IMRPhenomD approximant~\cite{PhysRevD.93.044007}, which is the GR model that is used in this paper. In Sec.~\ref{sec:BeyondGR}, we discuss waveform models in the context of modelling beyond-GR behavior, and the various ways in which inspiral models can be extended up to merger. In Sec.~\ref{sec:pe}, we discuss how our beyond GR model can be used in GW parameter estimation. In Sec.~\ref{sec:conclusions}, we summarize the results and lay out possible directions in which this work can be extended.

\section{GR Approximants\label{sec:grapprox}}
Deducing source properties from a detected GW signal from a compact binary coalescence requires the production of a candidate waveform, called a template, against which the strain data can be compared. In practice, parameter estimation requires the generation of millions of templates of differing parameters~\cite{PhysRevD.46.5236}. Analytically, we have the post-Newtonian (PN) theory~\cite{Blanchet2014-kc} which gives an accurate approximation to the inspiral GW for large separations and slow velocities. For close separations and high velocities before merger, we need full Numerical Relativity (NR) calculations. However, it is prohibitively expensive to have NR simulations for each choice of candidate parameters.  Depending on the degree of accuracy required, there are various alternatives. GR-based phenomenological waveform models are a computationally cheaper method used to more rapidly produce many candidate waveforms. The trade-off is that these GR approximants are calibrated such that they are only consistent with NR simulations in a limited region of parameter space. Moreover, NR itself encounters limits to its accuracy that are comparable to LIGO's accuracy needs~\cite{Okounkova2020-nf, Chu:2015kft, Boyle:2019kee}. To that extent, such waveforms can only be said to be GR waveforms up to a certain degree of accuracy; they are only GR approximants.

While there are approximants available, and in the process of being developed, that include higher-order modes~\cite{PhysRevD.103.104021, Cotesta:2020qhw, Garcia-Quiros:2020qpx}, in this paper we will restrict ourselves to the dominant $l=|m|=2$ modes only~\cite{PhysRevD.93.044006}.

\subsection{\label{sec:TaylorF2}TaylorF2}
For completeness, we reproduce here the TaylorF2 approximant~\cite{Blanchet2014-kc} that models the $\ell=|m|=2$ modes of GWs from binaries inspiraling on quasi-circular orbits. The TaylorF2 approximant is obtained by applying the stationary phase approximation to a PN treatment of the two-body problem that assumes large separations and non-relativistic speeds.
It is analytic with no free model parameters, meaning it exists independent of calibration to any NR waveforms. The leading order amplitude and phase of $\tilde{h}_{\mathrm{TF2}}(f)$, in geometrized units where $G=c=1$, can be written as:
\begin{align}
\tilde{h}_{\mathrm{TF2}}(f)&=\sqrt{\frac{2\eta}{3\pi^{1/3}}}f^{-7/6}e^{i\phi_{\mathrm{TF2}}},
\label{eq:TF2}\\
\phi_{\mathrm{TF2}}&=
2\pi ft_c-\phi_c-\pi/4 \notag\\
&\quad{}+\frac{3}{128\eta}(\pi Mf)^{-5/3}\sum_{i=0}^{7}\phi_i(\pi Mf)^{i/3},
\label{eq:IMRPhenomD_inspiral}
\end{align}
where overhead tilde denotes that the quantity has been Fourier transformed into the frequency domain. The above equations give the frequency-domain waveform for a binary with component masses $m_1$ and $m_2$, with total mass $M=m_1+m_2$ and symmetric mass ratio $\eta=m_1 m_2/M^2$, that coalesces at a time $t_c$ with a coalescence phase of $\phi_c$. Here $\phi_i$ are the coefficients corresponding to the PN expansion, where each coefficient depends on intrinsic binary parameters. We refer the reader to Eq.(318) in~\cite{Blanchet2014-kc} for the spin-independent $\phi_i$. The TaylorF2 portion of IMRPhenomD is supplemented with additional spin-dependent corrections, which are given in~\cite{PhysRevD.93.044007}. 

\subsection{\label{sec:IMRPhenomD}IMRPhenomD}
The waveform approximant to which we apply the ppE correction is the IMRPhenomD approximant described in~\cite{PhysRevD.93.044007}. This is a frequency-domain approximant to the $\tilde{h}_{22}(f)$ GW mode. The approximant is split into three frequency regimes, denoted as the inspiral, intermediate, and merger-ringdown regimes. The inspiral regime extends from the low-frequency limit of the waveform up to a frequency of $Mf = 0.018$ in geometrized units, while the intermediate region extends from this point up to $f = 0.5f_{\mathrm{RD}}$, where $f_{\mathrm{RD}}$ is the ringdown frequency of the final Kerr black hole resulting from the coalescence.

In frequency-domain approximants, it is common to factor the strain $\tilde{h}_{\mathrm{GR}}(f)$ into separate amplitude and phase components:
\begin{equation}
\tilde{h}(f) = 
                \mathcal{A}(f)e^{i\phi(f)}.
\label{eq:generic}
\end{equation}
In the IMRPhenomD approximant, both $\mathcal{A}(f)$ and $\phi(f)$ are given as piecewise functions, with each piece corresponding to a different frequency regime.

In the inspiral regime, the IMRPhenomD waveform is the same as that of the TaylorF2 approximant, including the terms corresponding to higher-order PN corrections in both the phase and amplitude. The full TaylorF2 phase used by IMRPhenomD is given by \eqref{eq:IMRPhenomD_inspiral}.

In the intermediate regime, the IMRPhenomD phase is given by
\begin{equation}
\eta\phi_{\mathrm{Int}} = \beta_0 +\beta_1 f + \beta_2 \log(f) - \frac{\beta_3}{3}f^{-3},
\label{eq:IMRPhenomD_intermediate}
\end{equation}
where $\beta_0$ and $\beta_1$ serve to ensure continuity and differentiability, and $\beta_2$ serves as a fitting coefficient used to reproduce NR waveforms to a desired tolerance.

In the merger-ringdown regime, the IMRPhenomD phase is given by
\begin{multline}
\eta\phi_{\mathrm{MR}} = \alpha_0 +\alpha_1 f - \alpha_2 f^{-1} + \frac{4}{3}\alpha_3f^{3/4} \\
+ \alpha_4\tan^{-1}\left(\frac{f-\alpha_5 f_{\mathrm{RD}}}{f_{\mathrm{damp}}}\right),
\label{eq:IMRPhenomD_mergerringdown}
\end{multline}
where $\alpha_0$, and $\alpha_1$ ensure continuity and differentiability of the phase across the two regimes. The remaining parameters $\alpha_2$, $\alpha_3$, $\alpha_4$, $\alpha_5$, $f_\mathrm{RD}$, and $f_\mathrm{damp}$ are determined via fits to NR simulations of GR mergers.

\section{\label{sec:BeyondGR}PPE Approximants}
Introduced by Yunes and Pretorius, the parameterized post-Einsteinian (ppE) formalism provides a way to modify frequency-domain GR waveforms according to the leading PN order deviation predicted by a beyond-GR theory~\cite{PhysRevD.80.122003}. The ppE framework consists of four parameters, $\alpha$, $\beta$, $a$, and $b$. The modification to the GR waveform is given by:
\begin{equation}
\tilde{h}_{\mathrm{ppE}}(f) = \tilde{h}_{\mathrm{GR}}(f)(1+\alpha u^a)e^{i\beta u^b}.
\label{eq:ppe}
\end{equation}

The parameters $\alpha$ and $\beta$ quantify the amplitude and phase deviation from $\tilde{h}_{\mathrm{GR}}$, while the parameters $a$ and $b$ are selected by the beyond-GR theory that is being considered. The quantity $u(f)=(\pi \mathcal{M} f )^{1/3}$ is proportional to the Keplerian velocity of the orbit,  where $\mathcal{M} = (m_1 m_2)^{3/5}/(m_1 + m_2)^{1/5}$ is the chirp mass of the binary.

The additional terms that appear in $\tilde{h}_{\mathrm{ppE}}(f)$ corresponding to $a$ and $b$ can be seen as corrections of a PN order relative to those that appear in $\tilde{h}_{\mathrm{GR}}(f)$. The relative PN orders of the correction terms that appear for a choice of $a$ and $b$ are $a/2$ and $(b+5)/2$, respectively. If we look at Eq.~\eqref{eq:ppe}, we can see that the exponent $a$ appears in the term with coefficient $\alpha$, which controls the deviation in the amplitude, while the exponent $b$ appears in the term with coefficient $\beta$, which controls the deviation in the phase. As the matched filtering process is much more sensitive to changes in the phase than in the amplitude, in this paper we will ignore the effects of the amplitude term, setting $\alpha$ to zero.

In Table~\ref{tab:table1}, we list a few examples of beyond-GR theories with their corresponding values of $a$ and $b$.
\begin{table}[b]%
\caption{\label{tab:table1}%
The values of $a$ and $b$ corresponding to each beyond-GR theory, along with the PN order of the phase correction.
}
\begin{ruledtabular}
\begin{tabular}{ccccccc}
\textrm{}&
$a$&

$b$&

PN&\\
\colrule
dCS & 4  & $-1$ & 2 \\
Einstein-\AE ther & 0 & $-5$ & 0 \\
EdGB & $-2$  & $-7$ & $-1$\\

\end{tabular}
\end{ruledtabular}
\end{table}

\subsection{\label{sec:ppE}Inspiral Correction}
The ppE parameters $\alpha$ and $\beta$ correspond to the leading-order PN corrections to GR predicted by the beyond-GR theory under consideration. The ppE corrections to higher order are unknown, so the ppE-corrected waveform can be expected to deviate from the true beyond-GR waveform at higher frequencies. Recently, numerical simulations have been successful in evolving binaries in some beyond-GR theories through merger~\cite{Okounkova2020-nf}. These numerical waveforms can be used to extend the ppE model to higher frequencies. The $\beta$-ppE correction is applied in the inspiral regime, which extends up to $Mf = 0.018$ in the IMRPhenomD model.

\subsection{\label{sec:ppE_ext}Post-inspiral-$\beta$ Corrections}
Besides changes to the phasing that might occur in the inspiral because of beyond-GR effects, an additional modification that might occur is a change in the merger time caused by a change in the inspiral rate. This raises the question of how we might extend the simple ppE model beyond the inspiral regime while still preserving the physical plausibility of the model. We examine a progression of increasingly sophisticated models to see how this merger-time modification naturally arises out of extending the inspiral correction in a manner that preserves continuity and differentiability. The difference between each of these models is determined by the choice of the ppE phase correction, denoted as $\Delta\phi_{\mathrm{ppE}}$. The amplitude $\tilde{h}_{\mathrm{GR}}(f)$ is left unchanged in each case.
The generic form of the ppE correction where the phase is modified is then given by:
\begin{equation}
\tilde{h}_{\mathrm{ppE}}(f) = 
                \tilde{h}_{\mathrm{GR}}(f)e^{i\Delta\phi_{\mathrm{ppE}}(f)}.
\label{eq:generic_ppE}
\end{equation}

We now consider various methods for handling the ppE phase correction beyond the inspiral regime.
\subsubsection{\label{sec:ppE_0}Zero Correction}
For completeness, we include the case here where we do not include a post-inspiral correction at all, and instead truncate the ppE waveform at a particular frequency $f_{\mathrm{IM}}$.
\begin{equation}
\tilde{h}^{\mathrm{Trunc}}_{\mathrm{ppE}}(f) = \begin{cases}
    \tilde{h}_{\mathrm{GR}}(f)e^{i\beta u^b}, & f < f_{\mathrm{IM}},\\
    0, & f_{\mathrm{IM}} \leq f.
\end{cases}
\label{eq:ppe_trunc}
\end{equation}
\subsubsection{\label{sec:ppE_C0}$C^0$ Correction}
\begin{figure}[ht]
\includegraphics[width=\columnwidth]{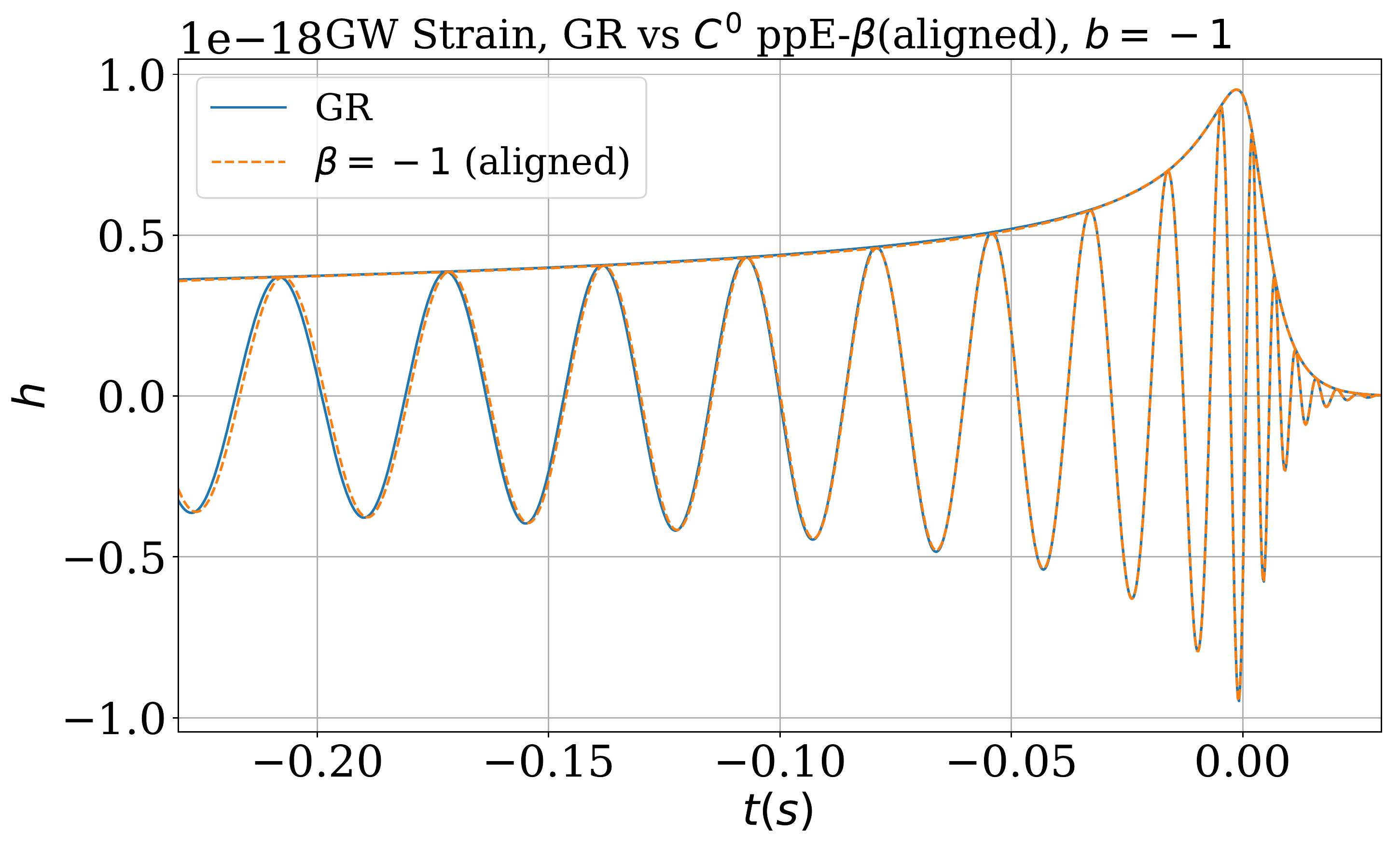}
\caption{\label{fig:grvcpe} A GR waveform with its ppE-$C^0$ counterpart. Despite being less physically motivated than other ppE schemes, the correction still results in a waveform that is not obviously unphysical. The mergers have been aligned in time and in phase. The amplitudes of the two waveforms are also plotted.}
\label{ppE_C0}
\end{figure}

Although the IMRPhenomD model is constructed to be $C^1$ in its phase, we include the $C^0$ correction here to be able to make comparisons against a naive post-inspiral correction that is not physically motivated.
\begin{equation}
\Delta\phi^{\mathrm{C0}}_{\mathrm{ppE}}(f) = \begin{cases}
  \beta u^b, &  f < f_{\mathrm{IM}},\\
  \beta u_{\mathrm{IM}}^b, & f_{\mathrm{IM}} \leq f.
\end{cases}
\label{eq:ppe_C0}
\end{equation}
As seen in Fig.~\ref{ppE_C0}, such a correction may not necessarily result in an obviously pathological waveform.

\subsubsection{\label{sec:ppE_C1}$C^1$ Correction}
\begin{figure}[ht]
\includegraphics[width=\columnwidth]{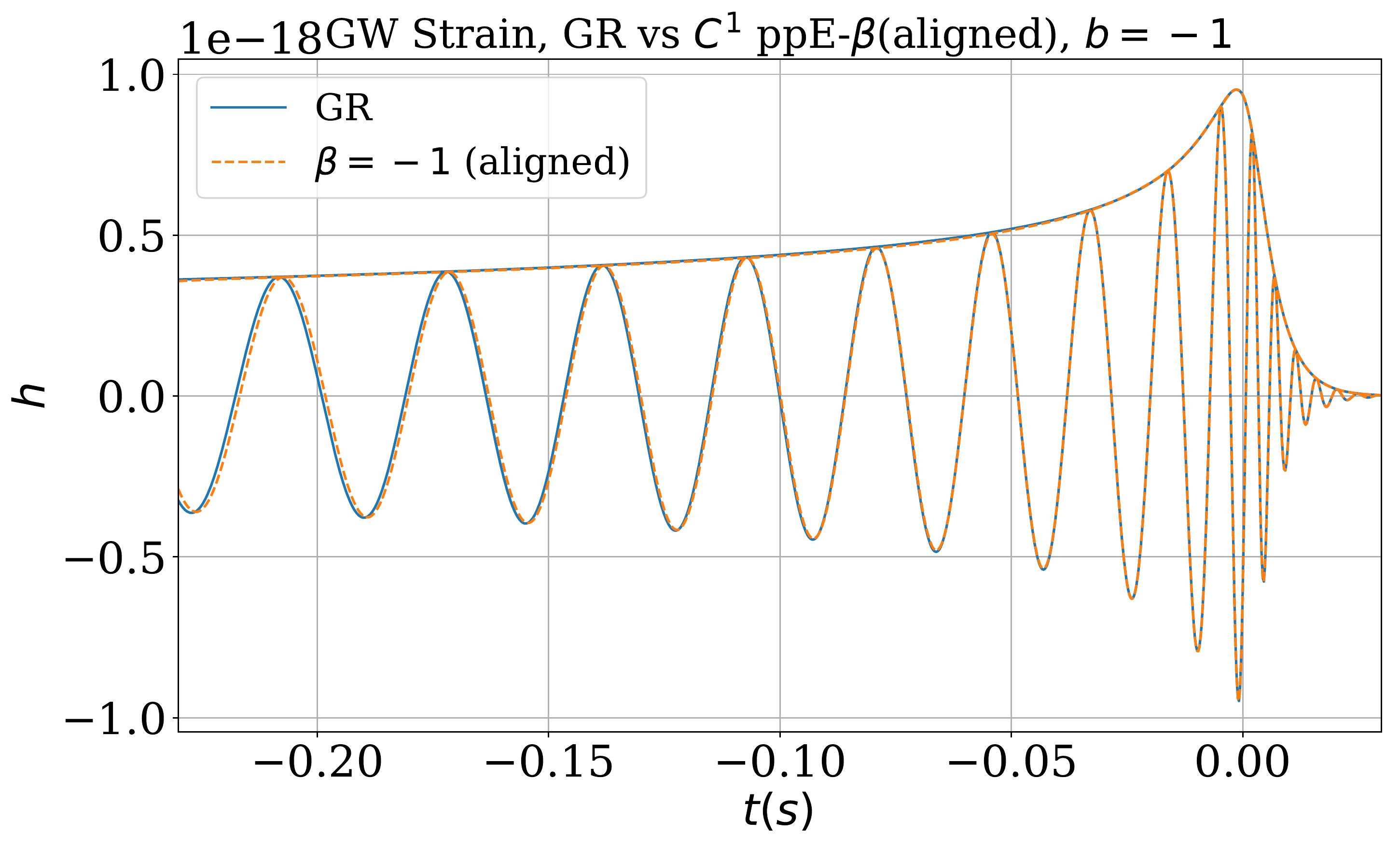}
\caption{\label{fig:grvcpe_beta} A GW in GR plotted against its ppE-$\beta$ counterpart in the time domain. The ppE waveform has been shifted in phase and time such that both waveforms have the same $t_c$ and $\phi_c$. In the frequency domain, aligning the waveforms in time renders them identical after $Mf_{\mathrm{IM}} = 0.018$, which is reached 0.05s before merger for this $M=80M_\odot$ binary. In the time domain, dephasing becomes apparent a few cycles before merger, but there is minimal amplitude disagreement between the two GWs.}
\label{ppE_C1}
\end{figure}

The $C^1$ correction is the first post-inspiral correction that is physically motivated.
\begin{multline}
\Delta\phi^{\mathrm{C1}}_{\mathrm{ppE}}(f) =  \\
\begin{cases}
  \beta u^b, & f < f_{\mathrm{IM}},\\
  \beta u_{\mathrm{IM}}^b  (1+\frac{b}{3}((u/u_{\mathrm{IM}})^3-1)), & f_{\mathrm{IM}} \leq f.
\end{cases}
\label{eq:ppe_C1}
\end{multline}
As seen in Fig.~\ref{ppE_C1}, the additional post-inspiral term induces a shift in the merger time and phase that leaves the two waveforms identical in the merger after this shift, which is linear in frequency, has been subtracted off. As the ppE-$\beta$ correction is only applied in the inspiral regime, this scheme is a good candidate for extending the correction into the merger/ringdown regime with minimal assumptions.

\subsubsection{\label{sec:ppE_CInf}$C^{\infty}$ Correction}
We also include the $C^{\infty}$ correction for the sake of completeness:
\begin{equation}
\tilde{h}^{C\infty}_{\mathrm{ppE}}(f) = 
                \tilde{h}_{\mathrm{GR}}(f)e^{i\beta u^b}.
\label{eq:ppe_Cinf}
\end{equation}
For some values of $b$ and $\beta$ the extension of the inspiral correction into the merger regime leads to significant distortion of the waveform. We compute an upper bound on the value of $\beta$ that is compatible with this type of correction. This motivates the need for a novel extension of the ppE correction into the intermediate and merger regimes.

\subsection{\label{sec:BeyondGR2}Post-inspiral-$\beta\epsilon$ Correction}
The $\beta$-ppE correction is extended into the intermediate and merger-ringdown regimes by demanding continuity and differentiability of the phase across frequency regimes.
This is equivalent to changing the parameters $\beta_0$, $\beta_1$, $\alpha_0$, and $\alpha_1$ in the IMRPhenomD approximant (See equations \eqref{eq:IMRPhenomD_intermediate} and \eqref{eq:IMRPhenomD_mergerringdown}). Consequently, the intermediate and merger-ringdown portion of the $\beta$-ppE corrected waveform is unchanged with respect to that of the original GR wavefrom from which it was constructed, albeit with a phase and time shift.
\subsubsection{\label{sec:MergerTimeCorrection}Merger Time Correction}
These shifts are given by:
\begin{align}
\phi_{\beta}& = \left(1-\frac{b}{3}\right)\beta u^b_{\mathrm{IM}}\\
t_{\beta} &= \frac{-b}{3}\frac{\beta u_{\mathrm{IM}}^b}{2\pi f_{\mathrm{IM}}}.
\label{eq:ppe_beta_shift_time}
\end{align}
This corresponds to the displacement of the peak GW amplitude in time and can be arranged such that at $t=t_c$, the GW reaches its peak amplitude in both the GR case as well as the ppE corrected case. The $\beta$-ppE correction used is the $C^1$ correction $\Delta \phi^{C1}_{\mathrm{ppE}}$ (see Eq.~\ref{eq:ppe_C1}).

\begin{figure}[ht]
\includegraphics[width=\columnwidth]{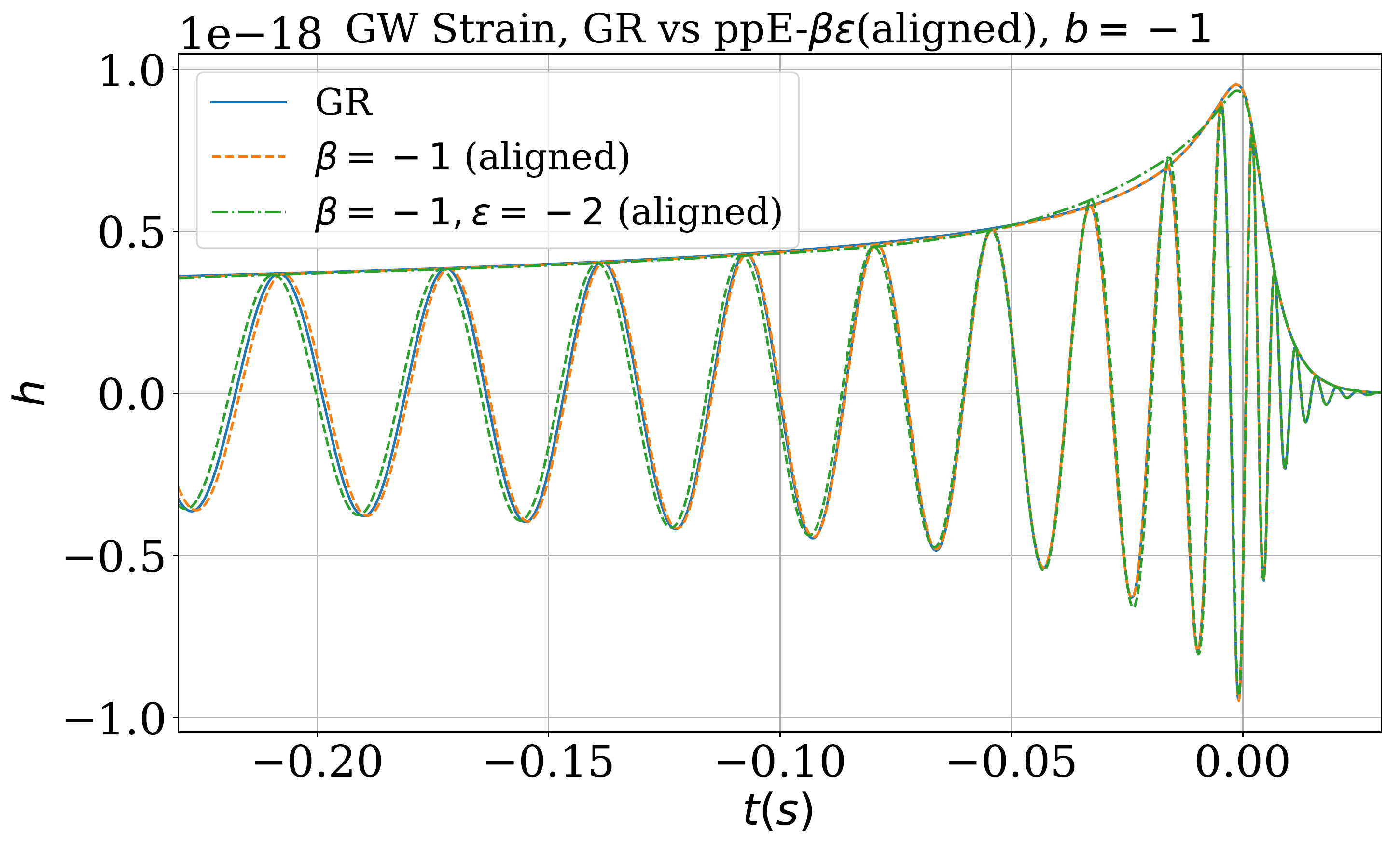}
\caption{\label{fig:grvcpe_betaepsilon} A GW in GR plotted against its ppE-$\beta$ and ppE-$\beta\epsilon$ counterpart in the time domain. The ppE waveforms have been shifted in phase and time such that all waveforms have the same $t_c$ and $\phi_c$. In the frequency domain, all waveforms are identical after $f = 0.5f_{\mathrm{RD}}$. In the time domain, dephasing becomes apparent a few cycles before merger, and the dephasing is more significant for the the ppE-$\beta\epsilon$ waveform. There is minimal amplitude disagreement between the three GWs.}
\label{ppE_C1_eps}
\end{figure}

In Fig.~\ref{ppE_C1_eps}, we show the effects of promoting the second of the $\beta$s, which affects the $u^3$ term, into an additional free parameter $\epsilon$ that controls the merger time.

\subsubsection{\label{sec:IMRPhenomDCompatibility}IMRPhenomD Compatibility}
\begin{figure}[ht]
\includegraphics[width=\columnwidth]{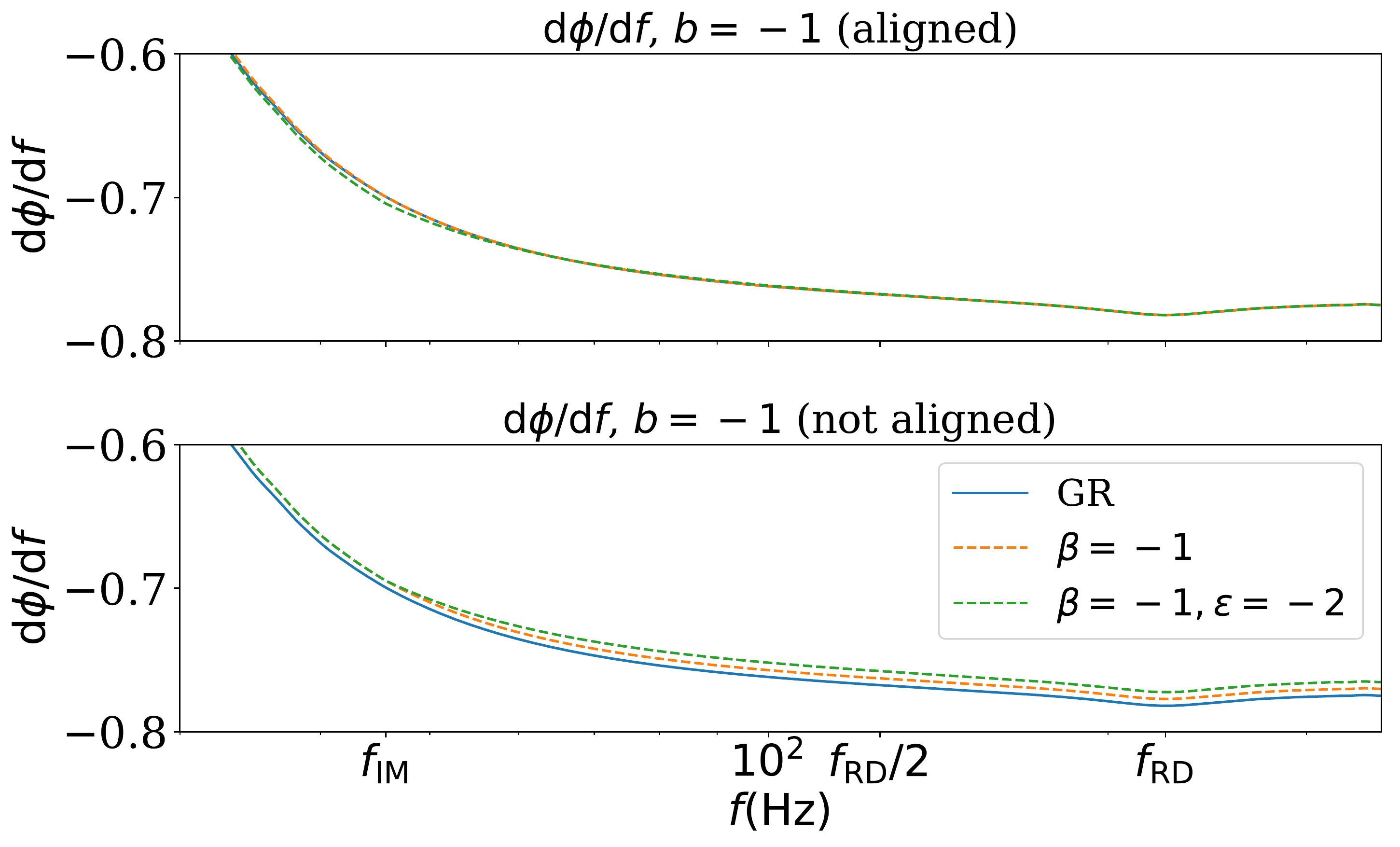}
\caption{\label{fig:grvcpe_dphidf} The derivative of the Fourier phase for both a GR waveform (blue), its ppE-$\beta$ counterpart (orange), and its ppE-$\beta\epsilon$ counterpart (green). The ppE-corrected waveforms are identical in the inspiral regime, as can be seen on the bottom panel. The ppE $\epsilon$ parameter controls the further deformation of the waveform beyond $Mf_{\mathrm{IM}} = 0.018$. The top panel illustrates how the three waveforms can have their mergers aligned in phase and time after applying the different corrections.}
\label{fourier_phase}
\end{figure}

The IMRPhenomD model uses a GR phase approximant that is C1 in frequency, so both the $\Delta\phi^{\mathrm{C0}}_{\mathrm{ppE}}$ and $\Delta\phi^{\mathrm{C1}}_{\mathrm{ppE}}$ post-inspiral phase corrections can be absorbed into a redefinition of the IMRPhenomD coefficients. This leaves the phase evolution $\Delta\phi'_{\mathrm{ppE}}(f)$ entirely expressible in terms of the IMRPhenomD parameters. This means that an IMRPhenomD waveform corresponding to a GR signal can be found such that the two frequency waveforms match exactly in the inspiral. In the C0 case, this corresponds to a change in the coalescence phase of the original GR waveform. In the C1 case, this corresponds to a change in both the coalescence phase and coalescence time of the original GR waveform. In Fig.~\ref{fourier_phase} we show how this $C^1$ agreement between the IMRPhenomD model and the ppE correction scheme is reflected in the fact that the derivatives of the Fourier phases are identical in the merger/ringdown regime, barring a constant offset corresponding to the time shift incurred by the ppE modification.

\subsubsection{\label{sec:TimeShiftingProperty}Time-Shifting Property}
Note that the time shift in the coalescence time incurred by extending the ppE-$\beta$ correction into the merger regime in a C1 fashion is not a prediction of the original beyond-GR theory in the ppE-$\beta$ framework. The ppE-$\beta$ correction is calculated using an assumption of a quasicircular inspiral. One should not expect that the dynamics of a merger in which this assumption breaks down can be approximated in this way. This motivates us to extend the ppE-$\beta$ correction by generalizing the time shift such that it is controlled by a new parameter $\epsilon$ (cf.~Eq.\eqref{eq:ppe_beta_shift_time}):
\begin{equation}
t_{\epsilon} = \frac{-b}{3}\frac{\epsilon u_{\mathrm{IM}}^b}{2\pi f_{\mathrm{IM}}}.
\label{eq:ppe_time_shift}
\end{equation}

\begin{figure}[ht]
\includegraphics[width=\columnwidth]{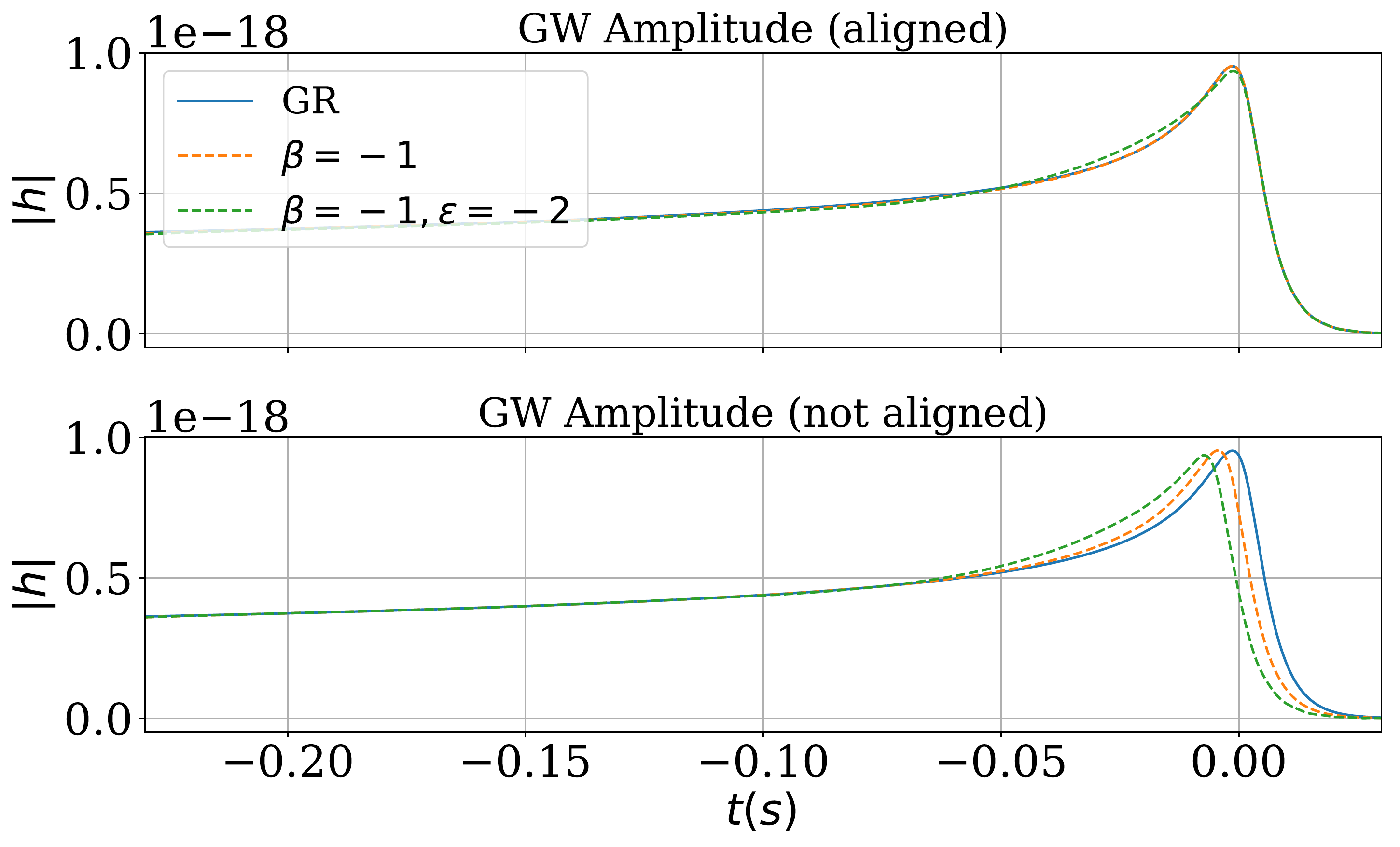}
\caption{\label{fig:grvcpe_eps_amps} The GW strain amplitudes after applying various ppE corrections, with the mergers aligned in time in the top panel. While the ppE-$\beta\epsilon$ correction is a pure phase correction in the frequency domain, this is not the case after transforming to the time domain. While the ppE-$\beta$ corrected GW suffers from less distortion, neither produce a GW amplitude that is non-monotonic before $t_c$.}
\label{gw_amplitude}
\end{figure}

\begin{figure}[ht]
\includegraphics[width=\columnwidth]{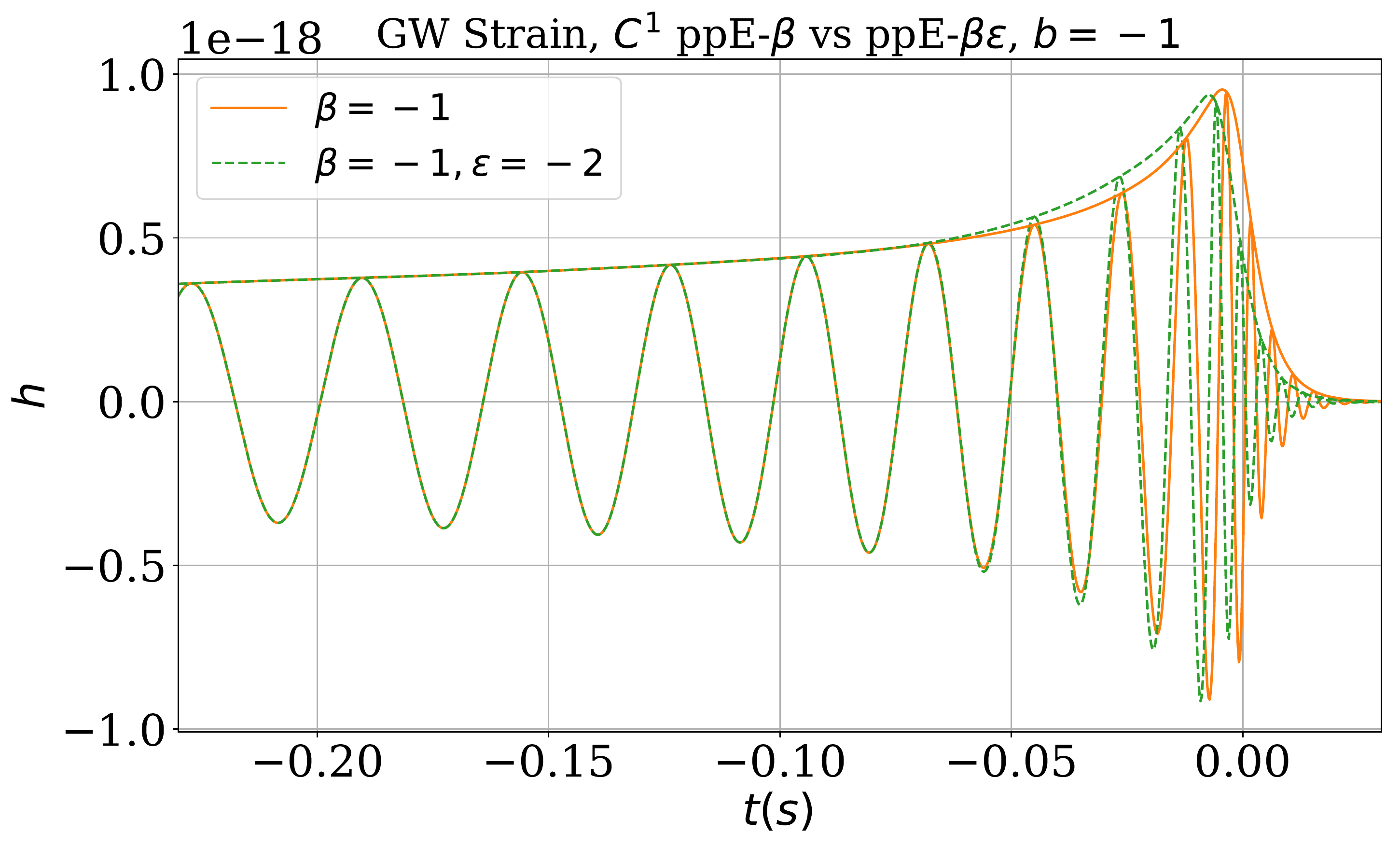}
\caption{\label{fig:grvcpe_epsilon} A ppE-$\beta$ corrected waveform plotted against its ppE-$\beta \epsilon$ corrected counterpart in the time domain. The ppE waveforms have \textbf{not} been shifted in phase and time such that all waveforms have the same $t_c$ and $\phi_c$. The localized shifting of the merger portion of the ppE-$\beta \epsilon$ corrected waveform relative to the ppE-$\beta$ corrected waveform can be clearly seen.}
\label{gw_beta_vs_eps}
\end{figure}

This new parameter $\epsilon$ then parameterizes the change in the merger time with respect to a GR merger with the same intrinsic parameters. Note that the time shift $t_{\beta}$ in the ppE-$\beta$ framework is determined by the C1 requirement imposed on the waveform across the frequency regimes. In the ppE-$\beta\epsilon$ framework, the C1 requirement is still imposed; the requirement is satisfied by modifying the waveform in the intermediate regime through changes in the parameters $\beta_0$, $\beta_1$, $\beta_2$, and $\beta_3$, which are now functions of $\epsilon$. The phase change in the inspiral-merger regime is then given by:
\begin{multline}
\Delta\phi^\epsilon_{\mathrm{ppE}}(f) = \\ \begin{cases}
  \beta u^b, & f < f_{\mathrm{IM}},\\
  \Delta\phi_{\mathrm{Int}}(f,\beta,\epsilon), & f_{\mathrm{IM}} \leq f < \frac{1}{2}f_{\mathrm{RD}},\\
  \beta u_{\mathrm{IM}}^b  +\frac{b}{3}\epsilon u_{\mathrm{IM}}^b((u/u_{\mathrm{IM}})^3-1), & \frac{1}{2}f_{\mathrm{RD}} \leq f.\\
\end{cases}
\label{eq:ppE_epsilon_phase_change}
\end{multline}

The form of $\Delta\phi_{\mathrm{Int}}(f,\beta,\epsilon)$ is determined by the coefficients $\beta_0$, $\beta_1$, $\beta_2$, and $\beta_3$ that solve the following matrix equation:

\begin{equation}
\begin{aligned}
\hspace{0pt} &\begin{bmatrix}
\Delta \phi^{\epsilon}_{\mathrm{ppE}}(f_{\mathrm{IM}})\\
\Delta \phi^{\epsilon \prime}_{\mathrm{ppE}}(f_{\mathrm{IM}})\\
\Delta \phi^{\epsilon}_{\mathrm{ppE}}(f_{\mathrm{RD}}/2)\\
\Delta \phi^{\epsilon \prime}_{\mathrm{ppE}}(f_{\mathrm{RD}}/2)
\end{bmatrix}=\\*
&\begin{bmatrix}
1 & f_{\mathrm{IM}} & \log((f_{\mathrm{IM}}) & -\frac{1}{3}f_{\mathrm{IM}}^{-3}\\
0 & 1 & 1/f_{\mathrm{IM}} & f_{\mathrm{IM}}^{-4}\\
1 & f_{\mathrm{RD}}/2 & \log(f_{\mathrm{RD}}/2) & -\frac{1}{3}(f_{\mathrm{RD}}/2)^{-3}\\
0 & 1 & 2/f_{\mathrm{RD}} & (f_{\mathrm{RD}}/2)^{-4}
\end{bmatrix}
\begin{bmatrix}
\beta_0\\
\beta_1\\
\beta_2\\
\beta_3
\end{bmatrix}.
\end{aligned}
\label{eq:matrix_eq}
\end{equation}

While Eq.~\eqref{eq:ppE_epsilon_phase_change} gives the modification of the
waveform, this is not the modification that is applied when performing
parameter estimation. The presence of the factor that is linear in frequency in
the ringdown regime causes the coalescence time to incur a shift after Fourier
transforming to the time domain; this linear factor is subtracted off in order
for the coalescence time $t_c$ to remain unchanged for different choices of
$\beta$ and $\epsilon$. Note that when a correction of non-zero $\beta$ is
applied to the inspiral regime in the $\beta\epsilon$ correction scheme, a
corresponding correction of $\epsilon = \beta$ must also be applied to the rest
of the waveform such that the resultant waveform is equivalent to a ppE-$\beta$
corrected waveform satisfying the C1 continuity condition. For example, in
Fig.~\ref{gw_beta_vs_eps}, a choice of $\epsilon = \beta = -2$ would lead to
the dotted orange curve lying on top of of the blue curve. A choice of
$\epsilon \neq \beta$ leads to the merger portion of the waveform being
displaced. As this leaves the rest of the waveform intact, it demonstrates how
$\epsilon$ is a nearly independent parameter from $\beta$, which only affects
the inspiral portion of the waveform, and thus can be measured separately. In
Fig.~\ref{gw_amplitude} the GW strain amplitudes are plotted demonstrating that
the monotonicity of the amplitude is preserved after the $\epsilon$
modification of the merger-ringdown portion of the waveform, indicating that
the time-shifting property of the $\epsilon$ modification is a physically
plausible one.

\section{Parameter Estimation\label{sec:pe}}
In order to determine the characteristics of the two compact objects that generate a detected GW, one finds the model parameters for a template waveform that best corresponds to the data. For GR approximants there may be up to fifteen such parameters, which includes extrinsic parameters determining the sky location of the binary as well as intrinsic parameters corresponding to binaries that are physically distinct in their source frame. The ppE-$\beta\epsilon$ corrected waveforms have two additional parameters, corresponding to beyond GR modifications in the inspiral and intermediate regimes. Constraints on these parameters can also be obtained using matched-filter parameter estimation, which allows us to begin examining the degree to which the current GW data deviates from GR, if at all.

\subsection{Match}
It may be the case that the GWs that observatories are detecting are beyond-GR in nature. That these observatories are still able to identify GW events in the data using GR templates tells us that the theory of gravity describing these GWs must not be too dissimilar from GR at these scales. To quantify the amount of mismatch between a GR waveform and a ppE-corrected waveform with non-zero $\beta$, we use the \textit{match}, defined as follows:
\begin{multline}
\mathrm{match}(h,h_{\mathrm{ppE}}) = \\ \max_{t_c, \phi_c} \Re \frac{\int df\hat{h}^*(f) \hat{h}_{\mathrm{ppE}}(f)e^{-2\pi ift_c+ i\phi_c}}{{\sqrt{\int df\hat{h}^*(f) \hat{h}(f)}\sqrt{ \int df \hat{h}_{\mathrm{ppE}}^*(f) \hat{h}_{\mathrm{ppE}}(f)}}} .
\end{multline}
where $\hat{h}(f)$ is the noise-weighted Fourier-domain waveform $\tilde{h}(f)/S_n(f)$, and $S_n(f)$ is the corresponding noise curve.

Using the ppE framework, we can produce waveforms that are distinguishable from GR. We then compare the similarity between the ppE modified waveform and the unmodified GR waveform using the match statistic between the two waveforms. This gives us a quantitative handle on how large a ppE correction we can make to the GR waveform.

\subsection{Injection Recovery}
In order to quantify the degree to which the beyond-GR parameters can be recovered from GW data, we carry out parameter estimation (PE) on simulated data. In general, this is done by choosing a waveform model to generate the signal to be injected in data that would be analyzed by a Bayesian parameter estimation pipeline. We then choose a second waveform model (which can be the same as the first) to generate the templates with which the matched-filtering is performed. We do not add a noise realization (i.e. we use {\it zero-noise}) which is equivalent to averaging over an ensemble of different noise realizations~\cite{Nissanke:2009kt}. We use the zero-detuning high-power Advanced LIGO noise curve for computing the Bayesian likelihood.

To establish a baseline for how well parameters can currently be determined from GW data, we show the results of a parameter estimation run using IMRPhenomD templates on IMRPhenomD injections. This stands in for the case where we assume GR is a sufficiently accurate model of gravity for the purposes of our current GW data. Next, we want to explore what happens if we use GR-approximants for templates when the underlying signal is beyond-GR in nature. We represent this possibility by showing the results of a PE run using IMR\-PhenomD templates on ppE-corrected IMRPhenomD injections. The posteriors can then be compared against the GR-to-GR case where the templates match the underlying theory. We expect a slight broadening and biasing of the posteriors as the GR model becomes a poorer approximation to the beyond-GR injections. We then show the case of a PE run where a ppE-corrected template is used against a ppE-corrected injection. In this case we expect to recover the underlying parameters of the binary to a similar accuracy as in the GR-to-GR case. Finally we show the case of PE with a ppE template against a GR injection, to represent the case where the new ppE model is used before GW detectors have the sensitivity to capture beyond-GR effects in GW data.

To simulate the effects of accumulating a large amount of GW signals that are beyond-GR in origin, we conduct the PE runs in the zero-noise limit. While an individual event may have a small SNR, having multiple detectors in place allows us to combine the detector data to allow the signals to add constructively, increasing the overall SNR. The zero-noise limit occurs in the limit of having infinitely many detectors for a single event~\cite{2014ApJ...784..119R}. Although one may also consider combining signals from multiple events, this must be done carefully as beyond-GR effects may manifest differently in binaries that have different parameters from each other.

\subsubsection{Parameter Estimation Algorithm}
The parameter estimation is done via a Markov Chain Monte Carlo (MCMC) algorithm within PyCBC Inference~\cite{Biwer2019-bo}, in which trial waveforms are sampled from a parameter space and then compared against the simulated data. Using the likelihood, subsequent waveforms are sampled in a probabilistic fashion, resulting in a chain of samples. If done correctly, the resulting distribution of samples approximates the posterior distribution. One detail of MCMC is that the sampling process resembles a random walk, and the samples that make up the initial steps of the random walk may not be representative of the posterior distribution, because of being initialized in, or stepping into, a region that should be exponentially less populated. These samples are identified and eliminated in a procedure known as \emph{burn-in}. There are multiple procedures for estimating when the chain has burned in; we use the \texttt{nacl} and \texttt{max\_posterior} methods defined in PyCBC. 

In our parameter estimation runs we use an extension of MCMC that makes use of what is known as \emph{parallel tempering}. Parallel tempering initializes multiple chains that explore the parameter space according to the ``temperature,'' which modifies the effect the likelihood has on the next iteration of the chain. This allows the chains to more efficiently explore the parameter space than could be done using only a single temperature.

The particular MCMC settings that have been used are given in table \ref{tab:MCMC}. 
\begin{table}[b]%
\caption{\label{tab:MCMC}%
The MCMC parameters used to perform the parameter estimation examined in this paper.
}
\begin{ruledtabular}
\begin{tabular}{cccccccc}
\textrm{$N_{\mathrm{walkers}}$}&
\textrm{$N_{\mathrm{temps}}$}&
\textrm{$N^\mathrm{Eff.}_{\mathrm{samples}}$}&
\textrm{$N^{\mathrm{Max}}_{\mathrm{per Chain}}$}&
\textrm{Burn-in Test}\\
\colrule
3000 & 4 & 3000 & 3000 & \texttt{nacl |  max\_posterior}\\

\end{tabular}
\end{ruledtabular}
\end{table}

\subsubsection{Choice of Prior Parameters}
In this section we discuss the effect the choice of prior has on the recovery of the parameters of a GR injection. At present, no measurable non-zero $\beta$-deviation from GR has been detected in GW data~\cite{Yamada2020-zu}. However, we can do better than to start with an arbitrarily wide uniform prior on $\beta$. From our match approximant, we can identify values of $\beta$ for which the match would drop as low as 95\%; we do not need to consider values of $\beta$ that would cause a higher mismatch as such deviations from GR would have been identified already~\cite{PhysRevD.96.084039}. For a fixed mismatch, we can also examine what effect the mass has on the value of $\beta$ needed to produce such a mismatch. At higher total masses, the lower frequency portions of the waveform, which are identically GR in the ppE-$\beta$ framework, dominate the SNR and a higher value for $\beta$ is needed for mismatches arising from the modified inspiral to be significant. This effect become more pronounced as one changes the value of the ppE parameter $b$ in the exponent of the phase change term. In this paper we are examining posteriors in $\beta$ and $\epsilon$ for only a single value of $b$, so even though the masses are not held fixed, a fixed prior over $\beta$ and $\epsilon$ is sufficient for our purposes.

The particular prior choices that have been used are given in table \ref{tab:prior}. 
\begin{table}[b]%
\caption{\label{tab:prior}%
The parameters of the priors used to perform the parameter estimation examined in this paper.
}
\begin{ruledtabular}
\begin{tabular}{c|cccccccc}
\textrm{param.}&
\textrm{$t_c$}&
\textrm{$M_1$}&
\textrm{$M_2$}&
\textrm{$\phi_c$}&
\textrm{$\beta$}&
\textrm{$\epsilon$}\\
\colrule
Dist. & Unif. & Unif. & Unif. & Unif. Angle & Unif. & Unif. \\
Min & 1126259462.32 & 10 & 10 & n/a & -20 & -20\\
Max & 1126259462.52 & 80 & 80 & n/a & 20 & 20\\

\end{tabular}
\end{ruledtabular}
\end{table}

\subsubsection{Choice of Injection Parameters}
In Table~\ref{tab:injection}, we list the different injections over which we have done parameter estimation runs. These values are the same as those used in the parameter estimation example in the PyCBC documentation. The masses probed are those most most likely to be found by detectors with noise curves similar to LIGO~\cite{PhysRevX.11.021053}. In this paper we only consider equal-mass injections, but the parameter estimation does not assume this.

\begin{table}[b]%
\caption{\label{tab:injection}%
The parameters used to generate the injections examined in this paper. Each of these entries corresponds to a set of 25 injections with $\beta$ and $\epsilon$ in the range of $[-5,5]\times[-5,5]$.
}
\begin{ruledtabular}
\begin{tabular}{c|cccccccc}
\textrm{param.}&
\textrm{ra}&
\textrm{dec}&
\textrm{$\iota$}&
\textrm{$\phi_c$}&
\textrm{$\psi$}&
\textrm{$d_L \mathrm{(Mpc)}$}&
\textrm{Mass 1}&
\textrm{Mass 2}\\
\colrule
set 1 & 2.2 & -1.25 & 2.5 & 1.5 & 1.75 & 100 & 30 & 30\\
set 2 & 2.2 & -1.25 & 2.5 & 1.5 & 1.75 & 100 & 35 & 35\\
set 3 & 2.2 & -1.25 & 2.5 & 1.5 & 1.75 & 100 & 40 & 40\\
set 4 & 2.2 & -1.25 & 2.5 & 1.5 & 1.75 & 100 & 45 & 45\\
set 5 & 2.2 & -1.25 & 2.5 & 1.5 & 1.75 & 100 & 50 & 50\\

\end{tabular}
\end{ruledtabular}
\end{table}

\subsubsection{GR Injection, GR Approximant}
\begin{figure}[ht]
\includegraphics[width=\columnwidth]{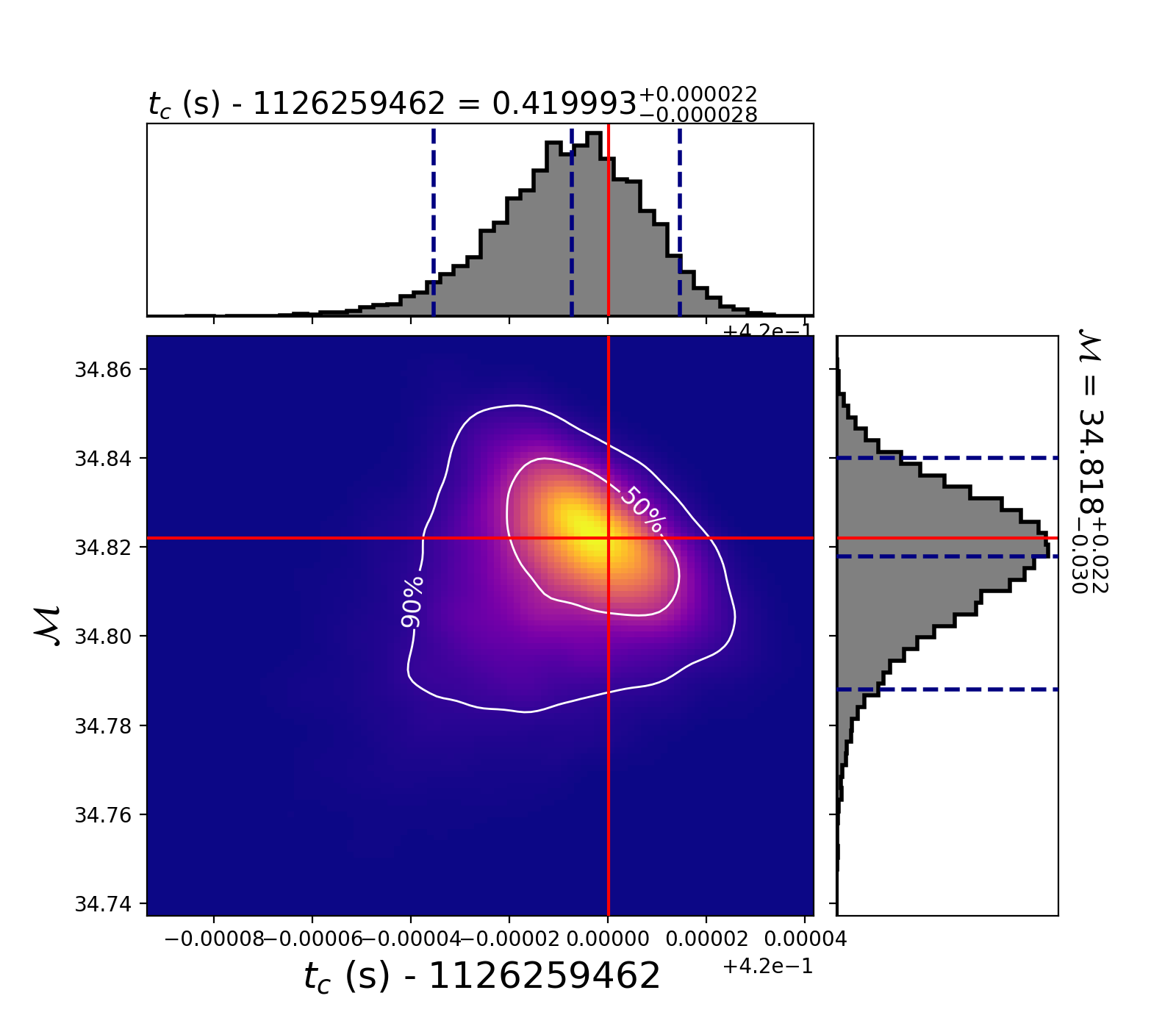}
\caption{\label{fig:gr_on_gr} A corner plot showing the result of using a GR approximant as the underlying template when the underlying signal is GR in nature. The red lines mark the values of the parameters of the injection that are to be recovered. In this case the the injection is an equal-mass binary with a total mass of 80 solar masses.}
\end{figure}

\begin{figure}[ht]
\includegraphics[width=0.9\columnwidth]{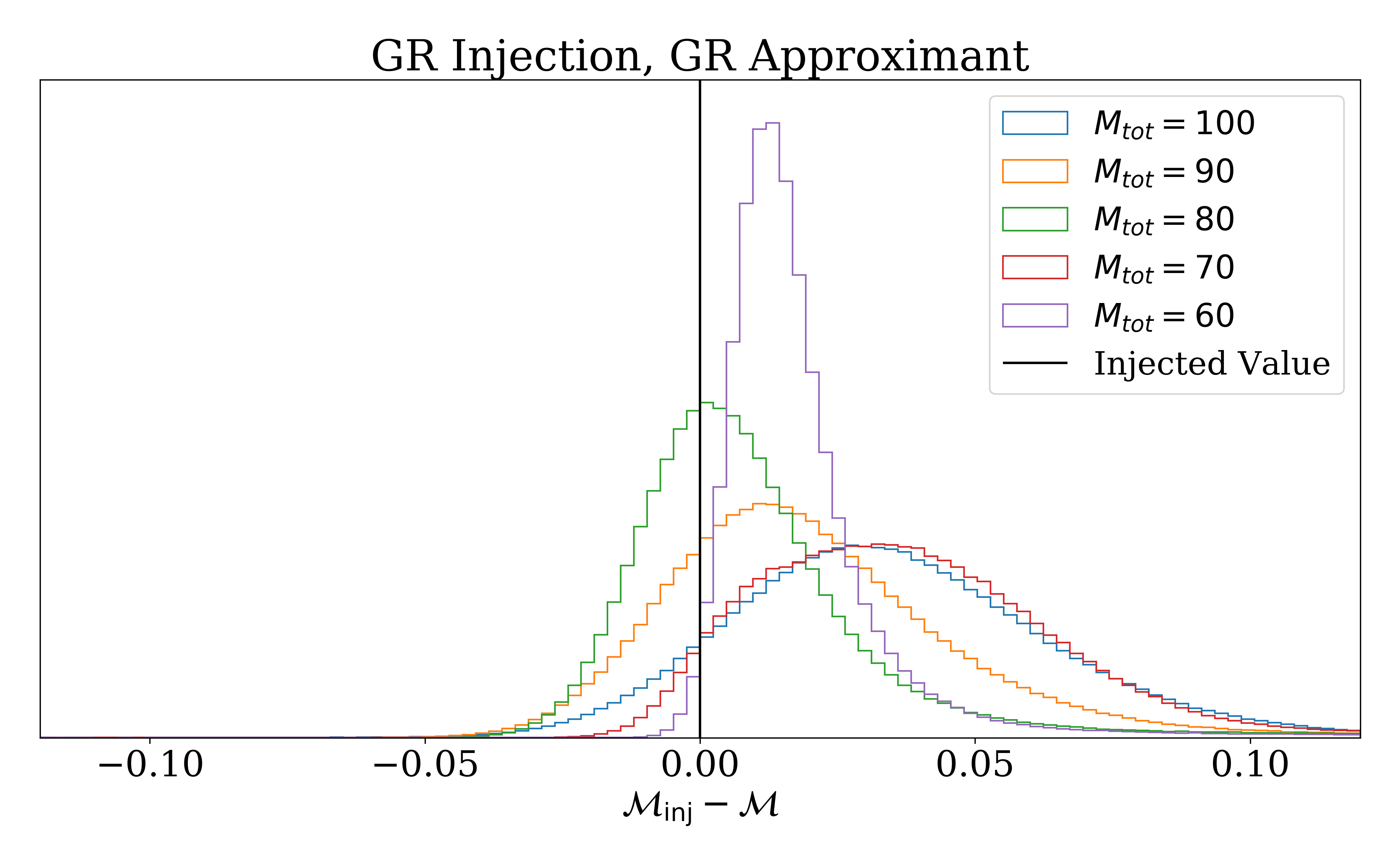}
\caption{\label{fig:gr_on_gr_chirpmass}Posterior distribution results of the chirp mass $\mathcal{M}$ from five separate parameter estimation runs, each of a system with a different total mass, plotted on top of one another. Each run uses an unmodified IMRPhenomD approximant waveform. The black vertical line indicates the point around which the posterior should ideally be centered, in the case where the chirp mass is recovered perfectly.}

\end{figure}

In order to have a baseline for how well we can expect the beyond-GR parameters to be recovered, we must first examine the degree to which the ordinary GR parameters can be recovered using the unmodified IMRPhenomD approximant. In Fig.~\ref{fig:gr_on_gr} we show the posterior recovered for two chosen parameters, the coalescence time $t_c$ and the chirp mass $\mathcal{M}$. In Fig.~\ref{fig:gr_on_gr_chirpmass}, we show the posterior distribution recovered for the chirp mass for each GR injection. Using the widths of the initial posteriors as a baseline, we repeat the parameter estimation on the GR injection using the ppE-$\beta$ approximant. If using the $\beta$-modified ppE approximant causes the spreads to widen too broadly, then parameter estimation may not be too useful anymore.

\subsubsection{GR Injection, ppE Approximant}
\begin{figure}[ht]
\includegraphics[width=\columnwidth]{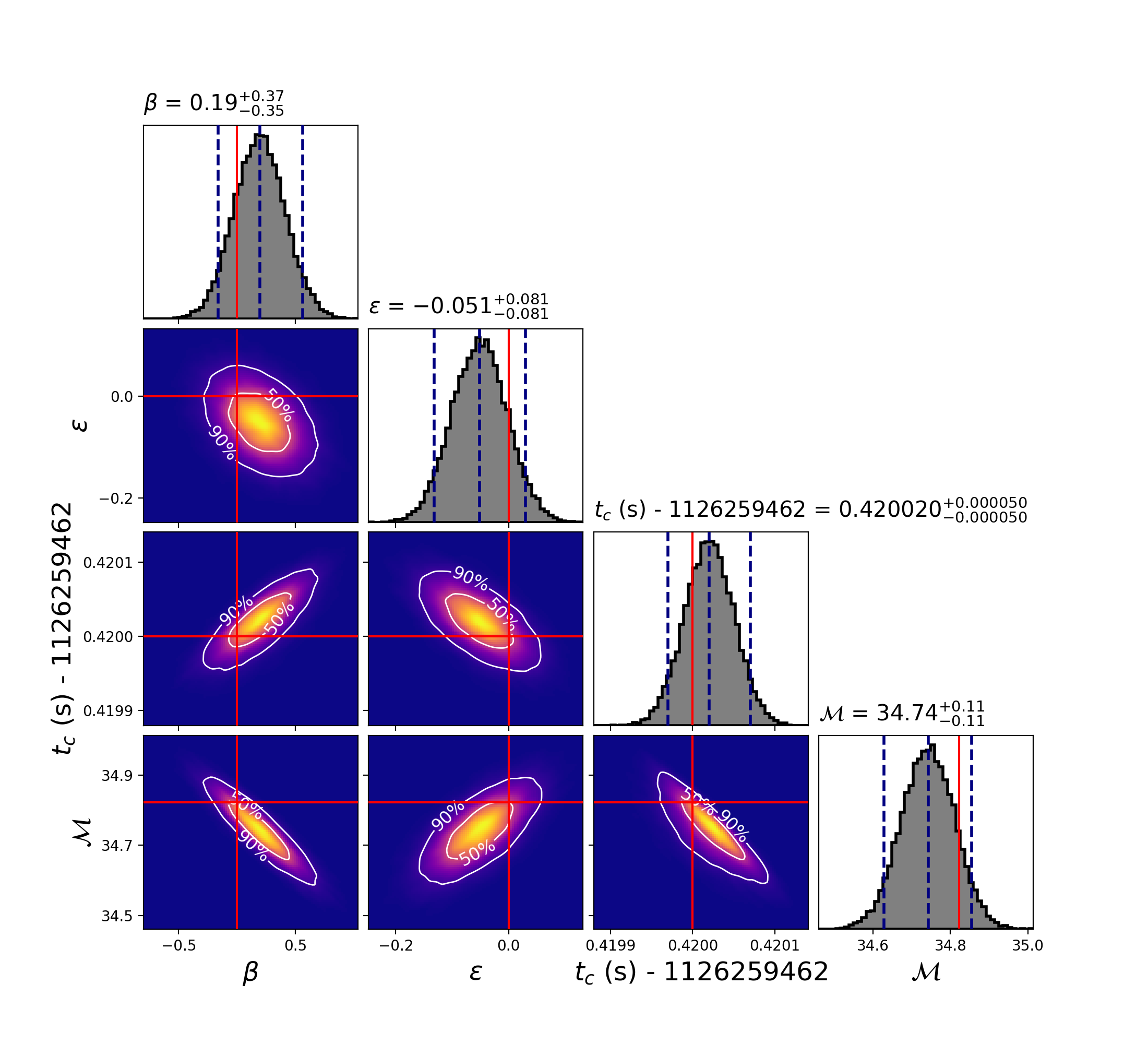}
\caption{\label{fig:gr_on_bgr} A corner plot showing the result of using a ppE-$\beta\epsilon$ approximant as the underlying template when the underlying signal is GR in nature. The red lines mark the values of the parameters of the injection that are to be recovered. Note that even in the case of a GR signal, which has $\beta=0$ and $\epsilon=0$, non-zero values are recovered. The injection is an equal-mass binary with a total mass of 80 solar masses.}
\end{figure}

\begin{figure}[ht]
\includegraphics[width=0.9\columnwidth]{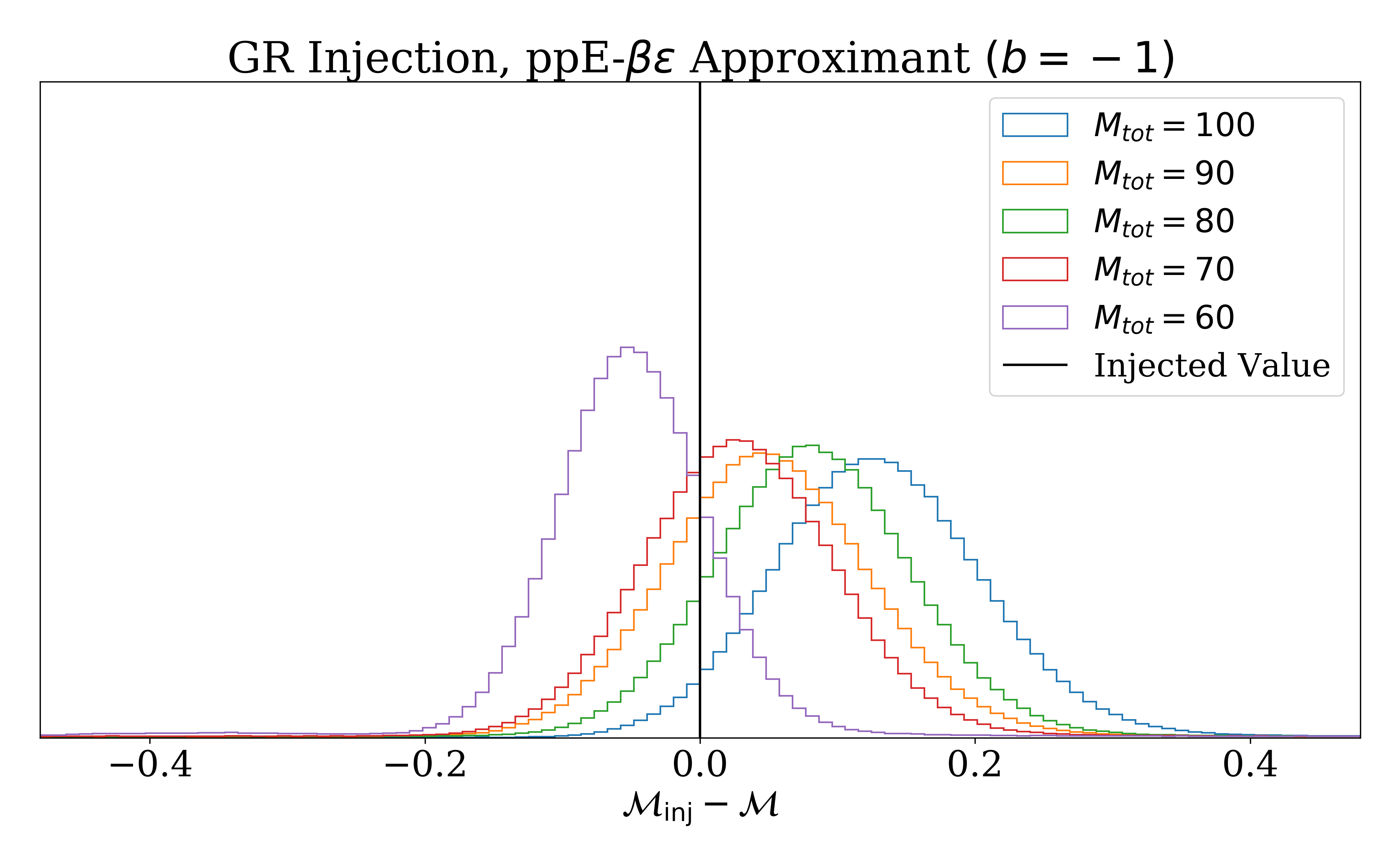}
\caption{\label{fig:gr_on_bgr_chirpmass}Posterior distribution results of the chirp mass $\mathcal{M}$ from five separate parameter estimation runs, each of a system with a different total mass, plotted on top of one another. Each run uses a ppE-$\beta\epsilon$ modified IMRPhenomD approximant waveform, where the ppE exponent $b$ is set to -1. The black vertical line indicates the point around which the posterior should ideally be centered, in the case where the chirp mass is recovered perfectly.}
\end{figure}

We now examine the results of parameter estimation runs where the injection is a GR waveform, but the template used to recover parameters is of a ppE nature. In this case, the parameter estimation will attempt to recover values of $\beta$ and $\epsilon$ in addition to the standard GR parameters. Since the underlying signal is GR in nature, we expect the recovered values of $\beta$ and $\epsilon$ to be consistent with zero. As we only vary the injections over the mass of the system, there are only five injections created. In Fig.~\ref{fig:gr_on_bgr}, we show the posteriors recovered for four chosen parameters, the coalescence time $t_c$ and the chirp mass $\mathcal{M}$ as before, and now the ppE parameters $\beta$ and $\epsilon$. We note that the peaks of the posteriors for $t_c$ and $\mathcal{M}$ are offset slightly from the posteriors recovered in Fig.~\ref{fig:gr_on_gr}. This offset reflects the change in the parameters recovered by the introduction of the additional parameters $\beta$ and $\epsilon$. Despite this offset, the injected value still lies within the 90\% contour lines. In Fig.~\ref{fig:gr_on_bgr_chirpmass} we show the corresponding plot to Fig.~\ref{fig:gr_on_gr_chirpmass}, which demonstrates how the posteriors for the chirp mass change after the introduction of $\beta$ and $\epsilon$ to the approximant. The injected value is still contained within each posterior across the masses probed, but the posteriors have also widened by a factor of roughly four compared to the GR approximant case.

\subsubsection{ppE Injection, GR Approximant}
\begin{figure}[ht]
\includegraphics[width=\columnwidth]{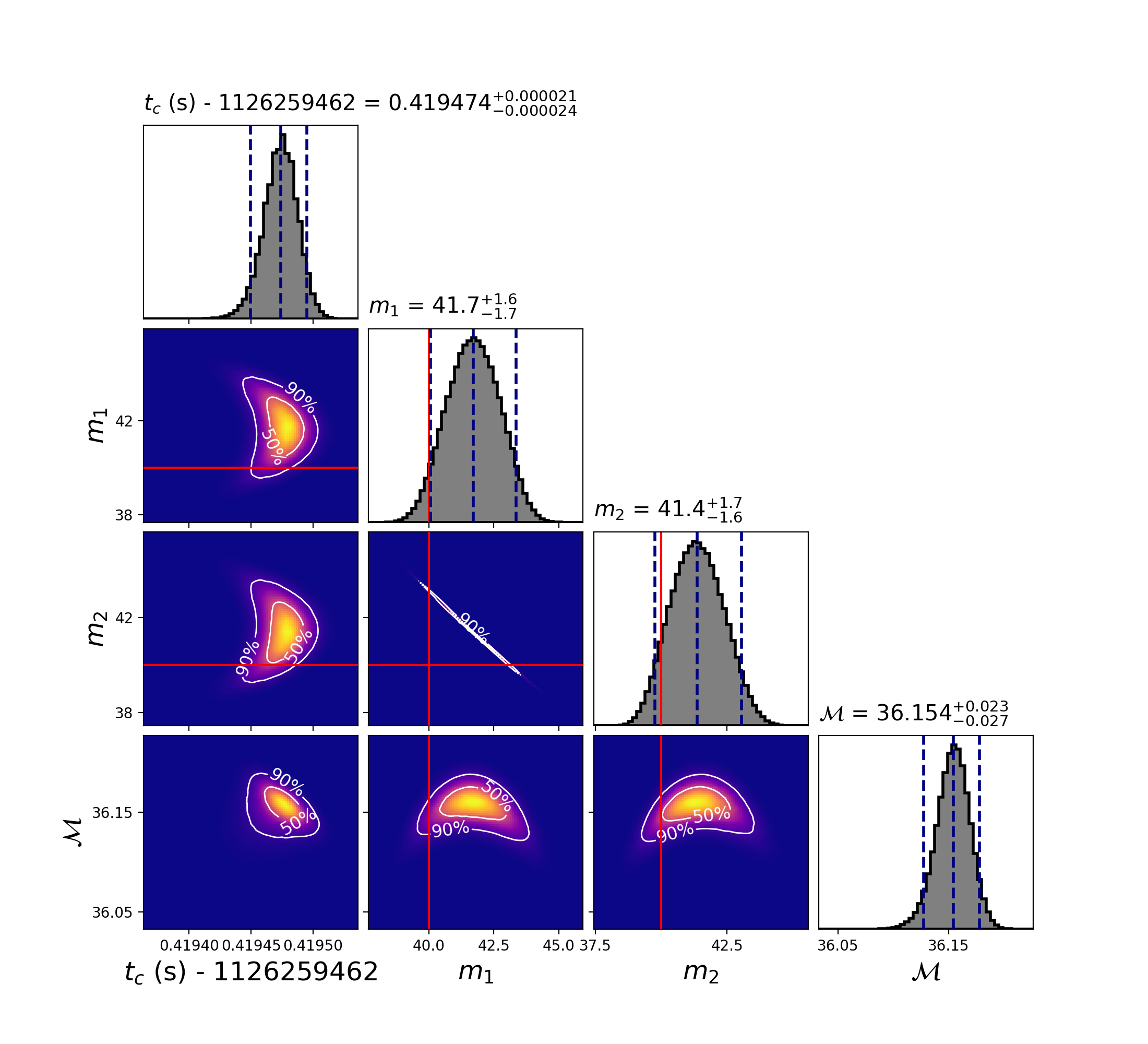}
\caption{\label{fig:71} A corner plot showing the result of using a GR approximant to model parameter estimation templates when the underlying signal is ppE-$\beta\epsilon$ in nature. The red lines mark the values of the parameters of the injection that are to be recovered. In this case the the injection is an equal-mass binary with a total mass of 80 solar masses, with $\beta = -3$ and $\epsilon$ = -1. The injected values for the chirp mass and the coalescence time are outside the domains of the plot, so their corresponding red lines are not visible.
}
\end{figure}

In this section, we examine the results of parameter estimation where the injection is a ppE waveform, while the template is of a GR nature. In this case, there are no $\beta$ or $\epsilon$ parameters in the template, and therefore there are no $\beta$ nor $\epsilon$ posteriors recovered. However, because the injections are of a ppE nature, they are created with varying $\beta$ and $\epsilon$ values. In this way, we can simulate how the presence of beyond-GR effects might instead be determined to be a GR signal, albeit one with slightly different parameters. In Fig.~\ref{fig:71} we can see an example of this; the posteriors recovered correspond to that of a more massive system.

\subsubsection{ppE Injection, ppE Approximant}
\begin{figure}[ht]
\includegraphics[width=\columnwidth]{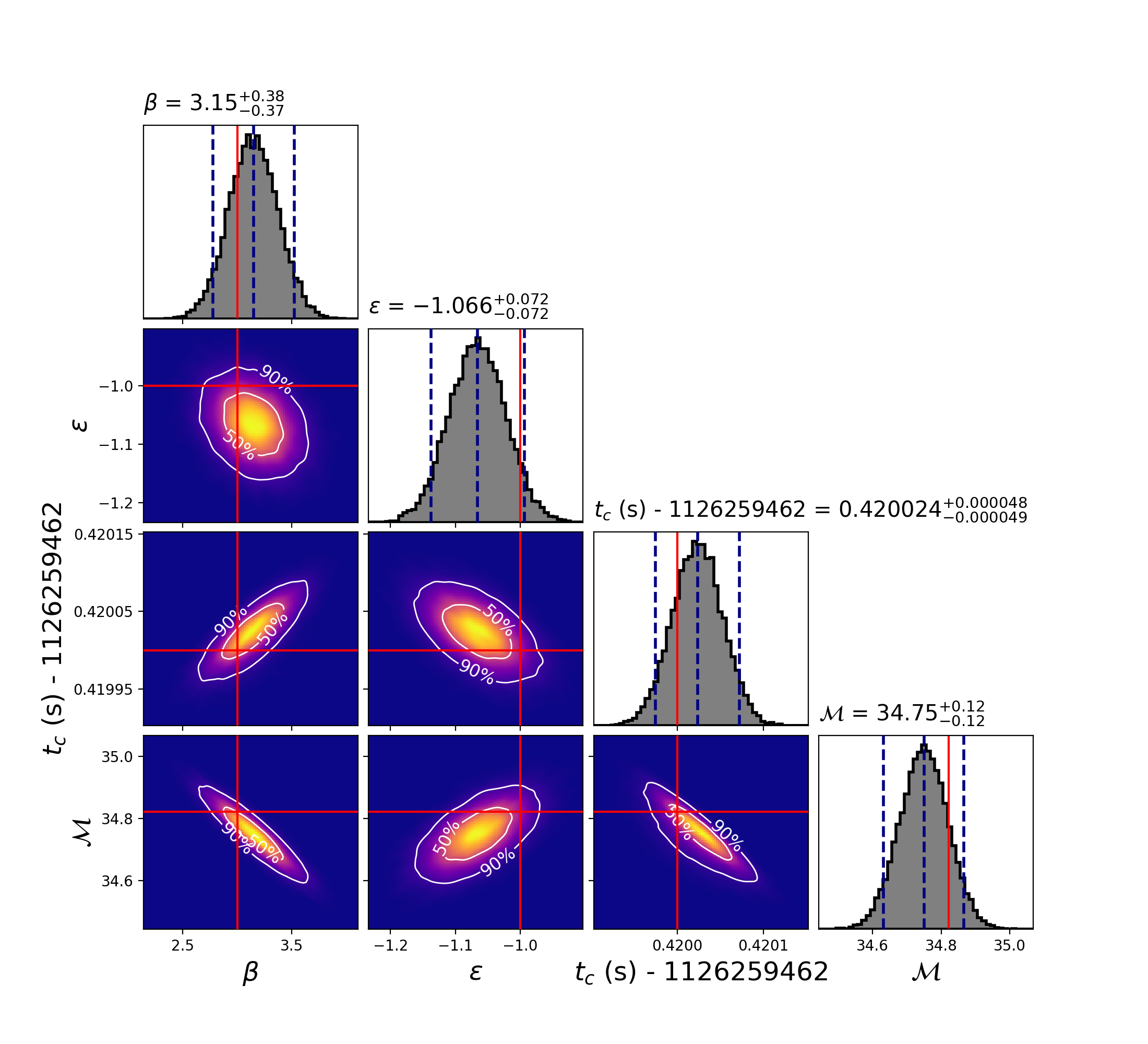}
\caption{\label{fig:bgr_on_bgr} A corner plot showing the result of using a ppE-$\beta\epsilon$ approximant as the underlying template when the underlying signal is also of a ppE-$\beta\epsilon$ nature. The red lines mark the values of the parameters of the injection that are to be recovered. We note that the posteriors recovered are not significanly different from the posteriors recovered in the GR injection, GR approximant case, and in the GR injection, ppE approximant case. As in the previous cases, the injection is an equal mass binary with a total mass of 80 solar masses, with $\beta = -3$ and $\epsilon$ = -1.}
\end{figure}

Finally, we examine the results of parameter estimation where both the injections as well as the templates are of a ppE nature. In this case, each injection is created with differing values of $\beta$ and $\epsilon$, and the templates will attempt to recover the standard GR parameters as well as $\beta$ and $\epsilon$ for each of these cases. In Fig.~\ref{fig:bgr_on_bgr} we see how both the $\beta$ and $\epsilon$ posteriors contain the injected value, indicating that both are simultaneously recoverable. The joint posterior distribution indicates that the correlation between $\beta$ and $\epsilon$ is less than or comparable to the correlation between either of these to the remaining parameters.

\subsubsection{ppE-$\beta \epsilon$ correction}
In the previous sections, we have examined whether the introduction of the $\beta$ and $\epsilon$ parameters spoils the recovery of GR parameters. To do this, we compared the posteriors for coalescence time $t_c$ and chirp mass $\mathcal{M}$ before and after this introduction, and found that the recovery is not spoiled.

In Figs.~\ref{fig:gr_on_gr_chirpmass} and \ref{fig:gr_on_bgr_chirpmass} we show this result is robust for the mass range probed by LIGO. where the injection is generated using IMRPhenomD and the template is the $\beta\epsilon$-corrected IMRPhenomD, in the limit of zero noise. 

Assuming the SNR is high enough to recover an accurate estimate of $\beta$, we can ask whether it is possible to recover the value of $\epsilon$ as well. One advantage of the $\epsilon$ extension to the ppE framework is that it is not necessary for $\beta$ to be recoverable for $\epsilon$ to be recovered as well. Depending on the total mass of the binary, a different part of the GW is in the LIGO band. Smaller mass binaries inspiral at higher frequencies, leaving more cycles in GW data. The longer inspirals aid in the recovery of $\beta$. On the other hand, if the GW is from a larger mass system, the power that is in the LIGO band is concentrated in the intermediate/merger-ringdown regime. As the $\epsilon$ parameter controls the deviation from GR in this region, it may still be recoverable in cases where $\beta$ is not. With multiple runs one can quantify how the width of the posterior behaves in the $\beta \epsilon$ parameter space.
When we plot the posterior distributions from different injections on top of one another, we see how sensitive the shape and location of the posteriors are on the beyond-GR parameters. To make clear how the parameters $\beta$ and $\epsilon$ control the waveform in different frequency regimes, we repeat the parameter estimation runs on injections of different total masses. As expected, when the total mass is smaller, is it the higher-frequency inspiral portion that lies within the LIGO band, and so the distinguishability of different values of $\beta$ is higher for lower system masses. Likewise, for higher mass systems, the lower-frequency merger falls within the LIGO band, and so the distinguishability of the different values of $\epsilon$ are higher here.

That $\beta$ and $\epsilon$ are independently recoverable is demonstrated best by posterior distributions like those shown in Fig.~\ref{fig:bgr_on_bgr}. Even as the late-inspiral parameter $\epsilon$ is changed, the resulting posterior distribution for $\beta$ appears to still contain the injected value. Demonstrating that this is the case is left to future work. While this relationship changes as the total mass of the system changes, depending on which parts of the inspiral fall into LIGO's sensitivity curve, it can be seen that both parameters are sufficiently distinguishable when both the inspiral and merger fall within LIGO's sensitivity curve.

\section{Conclusion\label{sec:conclusions}}
With tens of gravitational-wave detections in every observing run, LIGO and Virgo observatories provide an intimate window into highly dynamical gravity~\cite{LIGOScientific:2021djp}. This provides for a valuable opportunity for us to probe for any possible deviations from General Relativity that may manifest in GW signals. While a theory-specific approach remains both analytically and computationally prohibitive, Yunes \& Pretorius (YP) proposed a phenomenological approach to measure beyond-GR effects in GW signals~\cite{PhysRevD.80.122003}, called the parameterized post-Einsteinian framework.

In this paper we have presented a crucial extension that takes the ppE framework beyond the inspiral regime into the intermediate and merger-ringdown regimes. Our extended-PPE framework is implemented as a generalization of the IMRPhenomD approximant \cite{PhysRevD.93.044007} and invokes one additional parameter that corresponds to a change of coalescence rate during post-inspiral and up to the final merger time. It can be easily generalized to use any GR-based waveform model as a base though.

We demonstrate that the recovery of the additional parameters in the extended-ppE framework does not significantly alter the precision with which we can recover the posteriors for other parameters of the model, including the masses of individual components as well as inspiral ppE parameters.
This allows us to perform parameter estimation using ppE-corrected waveforms where only the inspiral correction is known. The additional parameter in the extended ppE model, $\epsilon$, controls the degree to which the inspiral correction (compared to GR based phasing) is extended into the intermediate and merger regimes.

For high values of the signal-to-noise ratio, deviations from GR, controlled by the ppE parameters $\beta$ and $\epsilon$, become significant enough to be measurable. Above a certain SNR, a value of zero for $\beta$ or $\epsilon$ falls outside the 95\% credible intervals for their posterior probability distributions if GR is not valid. The value of the ppE parameter $b$ for which this occurs will determine the class of beyond-GR theories that are most supported by GW data. We defer a systematic study of this to future work. In future work we propose to contrain both $\beta$ and $\epsilon$ using the latest catalog of LIGO-Virgo events.

\section*{Acknowledgements}
We thank \'Eamonn O'Shea and Masha Okounkova for useful discussions,
and Shasvath Kapadia and Daiki Watarai for their careful reading of the manuscript.
This work was supported in part by the Sherman Fairchild Foundation,
and by NSF Grant PHY-1912081 at Cornell. P.K. acknowledges support of
the Department of Atomic Energy, Government of India, under project
no. RTI4001; and by the Ashok and Gita Vaish Early Career Faculty
Fellowship at the International Centre for Theoretical Sciences.
The authors are grateful for computational resources provided by the
LIGO Laboratory and supported by National Science Foundation Grants
PHY-0757058 and PHY-0823459.

\bibliography{NewBibNov2}

\end{document}